\documentstyle[11pt,newpasp,twoside,epsf]{article}
\def\degr{^{\circ}}

\def\edcomment#1{\iffalse\marginpar{\raggedright\sl#1\/}\else\relax\fi}
\reversemarginpar

\begin{document}
\title{The large-scale Galactic magnetic field structure and pulsar
 rotation measures}
\author{J. L. Han$^1$, R. N. Manchester$^2$, G. J. Qiao$^3$ and A. G. Lyne$^4$
        }
\affil{$^1$ National Astronomical Observatories, CAS, Beijing 100012, China\\
$^2$ Australia Telescope National Facility, CSIRO, Epping,
         Australia\\
$^3$ Department of Astronomy, Peking University,
        Beijing 100871, China\\
$^4$ University of Manchester, Jodrell Bank Observatory, SK11 9DL,
         UK
      }

\begin{abstract}
Pulsars provide unique probes for the {\it large-scale} interstellar
magnetic field in the Galactic disk. The Parkes multibeam pulsar
survey has discovered many distant pulsars which enables us for the
first time to explore the magnetic field in most of the nearby half of
the Galactic disk. The fields are found to be {\it coherent in
direction} over a linear scale of $\sim 10$ kpc between the
Carina-Sagittarius and Crux-Scutum arms from $l\sim45\degr$ to
$l\sim305\degr$. The coherent spiral structures and field direction
reversals, including the newly determined counterclockwise field near
the Norma arm, are consistent with bi-symmetric spiral model for the
disk field. However, the antisymmetric rotation measure sky from the
Galactic halo and the dipole field in the Galactic center suggest that
the A0 dynamo is operating there.
\end{abstract}

\vspace{-5mm}
\section{Introduction}
\vspace{-2mm}
\begin{figure}[thb]
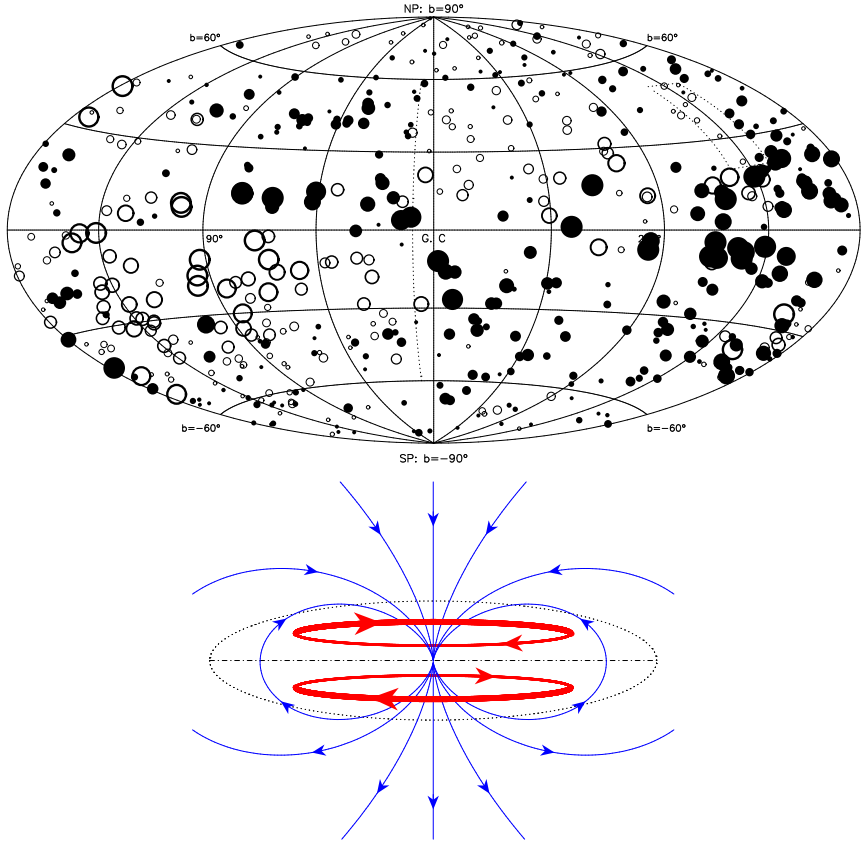

\plotfiddle{ers.ps}{60truemm}{270}{40}{40}{-190}{215}
\plotfiddle{a0_dynamo.ps}{40truemm}{270}{40}{40}{-150}{205}
\caption{The RMs of extragalactic radio sources show the antisymmetric
field structure of the Galactic halo, which is the same as that of an
A0 dynamo (see Han et al. 1997 for details).  Filled symbols denote
positive RMs and open symbols negative RMs. The distribution of
pulsar RMs is very similar but with generally smaller RM values.}
\end{figure}

Magnetic fields of celestial objects can be revealed by observations
of the Zeeman splitting of spectral lines, 
of polarized dust emission at mm, submm or infrared bands, 
of starlight polarization due to anisotropic scattering by magnetic-aligned 
dust grains, of radio synchrotron
emission and of Faraday rotation (see Han \& Wielebinski 2002). The first
two approaches have been used to detect, respectively, the radial
strength and transverse
orientation of magnetic fields in the molecular clouds, 
with a scale length of a few pc to hundreds pc. The starlight polarization
can only be used to delineate the orientation of transverse magnetic field
in the interstellar medium in the vicinity of the Sun within 2 or 3 kpc.
The polarized synchrotron emission, if observed from nearby spiral
galaxies, directly indicates the orientation of ordered magnetic
field; if observed from our own Galaxy,
it shows the large-angular-scale features emerging from the Galactic disk
due to the transverse magnetic fields, such as the North Polar Spur or
the vertical filaments near the Galactic center. The Faraday rotation of
pulsars and extragalactic radio sources are in fact the best probes
of the Galactic magnetic field on the large scale!

The rotation measure (RM) of a source is proportional to the
integration along its line of sight of the product of the radial
magnetic field and electron density, $RM=\int_{\rm obs}^{\rm obj}
B_{||} n_e dl$. Marvellously our Galaxy is the visually largest edge-on
Galaxy covering all the sky. There are so many sources, either the
extragalactic polarized radio sources or pulsars in our Galaxy, in all
directions that contain information of the field $B_{||}$ in all parts
of the Galaxy. This gives us a unique chance to study the detailed
structure of the magnetic field in the Galactic halo and disk, with a
principle similar to the CT technology used medically. We emphasize
that this is not possible for any other galaxies. The antisymmetric RM
distribution of extragalactic radio sources as well as pulsars,
covering all the inner Galaxy, reflects the azimuthal field with
reversed directions above and below the Galactic plane (Fig.1),
entirely consistent with the field configuration of an A0 dynamo,
whose origin probably lies in a dynamo in the halo (Han et al. 1997).
The polar field in the Galactic center strongly supports this
conclusion.

\vspace{-5mm}
\section{Pulsar rotation measures and the Galactic magnetic fields}
\vspace{-2mm}
\begin{figure}[thb]
\plotfiddle{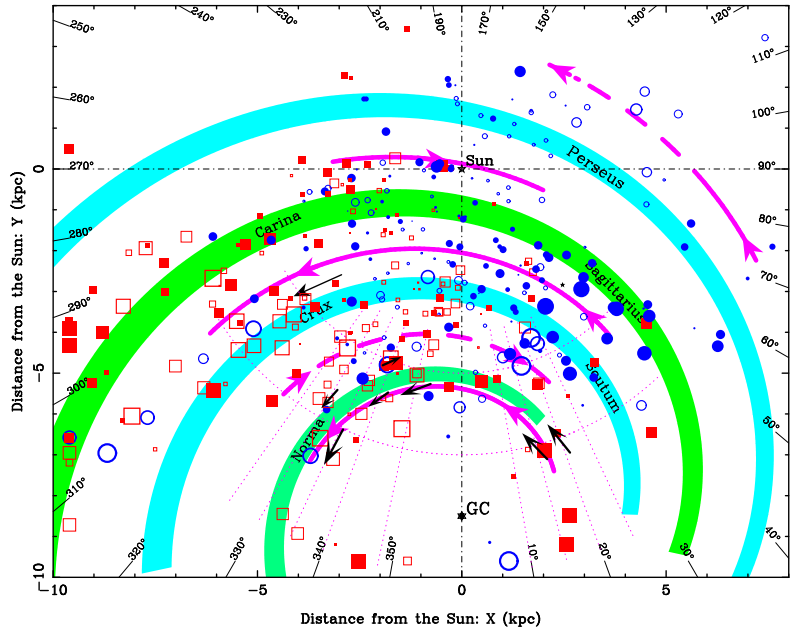}{79truemm}{270}{48}{48}{-183}{275}
\caption{The distribution of pulsar RMs projected onto the Galactic plane
reveals the field structure in the Galactic disk, which has
direction reversals from arm to arm (after Han et al. 2002). The square symbols
are newly determined RMs
from Parkes observations, and filled symbols indicate positive RMs.
The well-determined field structure is illustrated by thick lines and arrows.
The thick dashed lines indicate the controversial field structure
which needs further data for confirmation. The newly determined
field directions near the Norma arm and the Crux arm are indicated
by thin arrows.}
 \end{figure}

Soon after the pulsars were discovered, Lyne \& Smith (1969) detected
the linear polarization of pulsars. They noted that this ``opens up
the possibility of measuring the Faraday rotation in the interstellar
medium'' which ``gives a very direct measure of the interstellar
magnetic field'', because the $\int_{\rm obs}^{\rm obj} n_e dl $ can
be measured by pulse dispersion measure $DM$ so that $\langle
B_{||}\rangle$ can be directly obtained from the ratio
$RM/DM$. Manchester (1972, 1974) first systematically measured a
number of pulsar RMs for Galactic magnetic fields and concluded that
the local uniform field is directed toward about $l=90\degr$.  Thomson \&
Nelson (1980) modelled the pulsar RMs and found the {\it
first} field reversal near the Carina-Sagittarius arm. The largest
pulsar RM dataset was published by Hamilton \& Lyne (1987). Then Lyne
\& Smith (1989) used pulsar RMs to further study the
Galactic magnetic field. They confirmed the first field reversal in
the inner Galaxy and found evidence for the field reversal in the
outer Galaxy by a comparison of pulsar RMs with those of extragalactic
radio sources. Rand \& Lyne (1994) observed more RMs of distant
pulsars and found evidence for the clock-wise field near the
Crux-Scutum arm.

Han et al. (1997) first noticed that the RM distribution of high latitude
pulsars is dominated by the halo azimuthal field (Fig.1). Afterwards,
any analysis of pulsar RMs for the disk field was confined to the pulsars
at lower Galactic latitudes. Han et al. (1999) then observed a good
number of pulsar RMs and divided all known pulsar RMs into those lying
within higher and
lower latitude ranges for studies of the halo and disk field, respectively. 
Up to now, among about 1385 known pulsars, 523 pulsars have measured
values of RM observed and 365 of them lie at lower latitudes ($|b|<8\degr$).

One should note three important factors in the diagnosis of the
large-scale field structure from pulsar RMs. First, one normally
assumes that the azimuthal field component $B_{\phi}$ is greater than
vertical and radial components $B_z$ or $B_r$.  This is reasonable and
has been justified (Han \& Qiao 1994, Han et al.  1999). Second, it is
the gradient of the average or general tendency of RM variations
versus pulsar DMs that traces the large-scale field.  The scatter of
the data about this general tendency is probably mostly due to the
effect of smaller scale interstellar structure, such as the HII
regions.  Finally, the large-scale field
structure should give coherence in the gradients for many adjacent but
independent lines of sight. This can be recognized from the pulsar RM
distributions (e.g. $l=\pm20\degr$ near the Norma arm in Fig.2).

Very recently, the Parkes multibeam pulsar survey (Manchester et
al. 2001) has nearly doubled the number of known pulsars. Many of the new
pulsars are distant and are located near the center of the
Galaxy. These pulsars are unique probes for large-scale magnetic
fields. Han et al. (2002) recently report the detection of
counterclockwise magnetic fields near the Norma arm, which are {\it
coherent in field direction} over more than 5 kpc along the arm. Our
present understanding of the structure of the {\it large-scale} magnetic
field in the Galactic disk is summarised in Fig.2, which is almost all
revealed by pulsar RMs.

\vspace{-5mm}
\section{Conclusions}
\vspace{-2mm} We conclude that based on all available information, the
global field structure is best described by two distinct components:
a bi-symmetric spiral field in the disk  with reversed direction from 
arm to arm, and an azimuthal field in the halo with reversed directions
below and above the Galactic plane.  The latter, together with  
a probably dipole field in the Galactic center, corresponds to an A0
dynamo field configuration. Pulsars are the most important probes for
detection of the {\it large-scale} interstellar fields.

\small
\acknowledgments
JinLin Han is supported by the National Natural Science
Foundation of China (19903003 and 10025313), the National
Key Basic Research Science Foundation of China (G19990754)
and the partner group of MPIfR at NAOC.

\end{document}